\documentclass[letter]{aa}
\usepackage{graphicx}
\usepackage{txfonts}

\def\Mjup{\hbox{$M_{\rm Jup}$}}
\def\Rjup{\hbox{$R_{\rm Jup}$}}
\def\aRs{\hbox{$a/R_\star$}}
\def\RpRs{\hbox{$R_{\rm p}/R_\star$}}
\def\Tmid{\hbox{$T_{\rm mid}$}}

\def\bjdtdb{\mbox{$\mathrm{BJD}_\mathrm{TDB}$}}

\usepackage[colorlinks=true,citecolor=blue]{hyperref}
\usepackage{natbib}
\begin{document}

   \title{The GTC exoplanet transit spectroscopy survey. VII.
         \thanks{Based on observations made with the Gran Telescopio Canarias (GTC), 
         at the Spanish Observatorio del Roque de los Muchachos of the Instituto de 
         Astrof\'{i}sica de Canarias, on the island of La Palma}}

   \subtitle{Detection of sodium in WASP-52b's cloudy atmosphere}

   \author{G. Chen\inst{1,2,3}
          \and
          E. Pall\'{e}\inst{1,2}
          \and
          L. Nortmann\inst{1,2}
          \and
          F. Murgas\inst{1,2}
          \and
          H. Parviainen\inst{1,2}
          \and
          G. Nowak\inst{1,2}
          }

   \institute{Instituto de Astrof\'{i}sica de Canarias, V\'{i}a L\'{a}ctea s/n, E-38205 La Laguna, Tenerife, Spain\\
         \email{gchen@iac.es}
         \and
             Departamento de Astrof\'{i}sica, Universidad de La Laguna, Spain
         \and
             Key Laboratory of Planetary Sciences, Purple Mountain Observatory, Chinese Academy of Sciences, Nanjing 210008, China
             }

   \date{Received March 6, 2017; accepted March 17, 2017}

 
  \abstract
  {We report the first detection of sodium absorption in the atmosphere of the hot Jupiter WASP-52b. We observed one transit of WASP-52b with the low-resolution Optical System for Imaging and low-Intermediate-Resolution Integrated Spectroscopy (OSIRIS) at the 10.4~m Gran Telescopio Canarias (GTC). The resulting transmission spectrum, covering the wavelength range from 522~nm to 903~nm, is flat and featureless, except for the significant narrow absorption signature at the sodium doublet, which can be explained by an atmosphere in solar composition with clouds at 1~mbar. A cloud-free atmosphere is stringently ruled out. By assessing the absorption depths of sodium in various bin widths, we find that temperature increases towards lower atmospheric pressure levels, with a positive temperature gradient of $0.88\pm 0.65$~K\,km$^{-1}$, possibly indicative of upper atmospheric heating and a temperature inversion. }


   \maketitle
%

\section{Introduction}\label{sec:intro}

The sodium (Na) and potassium (K) doublets are two of the most important opacity sources for exoplanet atmospheric characterization in the optical wavelengths \citep{2000ApJ...537..916S}, and have been detected in around ten hot giant planets spanning a temperature range of 960--1740~K \citep[e.g.,][]{2002ApJ...568..377C,2008ApJ...673L..87R,2011MNRAS.412.2376W,2012MNRAS.426.1663S,2016Natur.529...59S,2017arXiv170200448W}. Recent ground-based observations have shown the great potential in robust detection and preliminary line-profile diagnosis for Na and K \citep[e.g.,][]{2012MNRAS.426.1663S,2015A&A...577A..62W,2015ApJ...803L...9H,2016ApJ...832..191N}. 

Here we present \object{WASP-52b} as another robust detection to this Na-K sample from our ground-based exoplanet transit spectroscopy survey \citep[e.g.,][]{2014A&A...563A..41M,2016A&A...585A.114P,2016A&A...594A..65N} with the 10.4~m Gran Telescopio Canarias (GTC). \object{WASP-52b} was discovered to transit a K2 dwarf every 1.75 days \citep{2013A_A...549A.134H}. This low-density planet \citep[0.43~\Mjup, 1.25~\Rjup;][]{2017MNRAS.465..843M} is highly inflated, and thus well-suited for transmission spectroscopy. The host star is active, and exhibits a rotational modulation of $16.4\pm0.04$ days with an amplitude of 9.6~mmag. Activity signatures such as chromospheric emission in the \ion{Ca}{ii} H+K lines \citep{2013A_A...549A.134H}, occulted star spots \citep{2017MNRAS.465..843M} and faculae \citep{2016MNRAS.463.2922K} have been identified. \citet{2016MNRAS.463.2922K} obtained a broadband transmission spectrum in the $u'$, $g'$ and a 31.2~nm filter centered on Na using the 4.2~m William Herschel Telescope, and found an increasing planetary radius towards the red optical. 

This paper is organized as follows. In Sect.~\ref{sec:data}, we summarize the observation and data reduction. In Sect.~\ref{sec:analysis}, we present the light-curve analysis and derive the absorption depths at Na and K. In Sect.~\ref{sec:discuss}, we discuss the atmospheric properties as indicated by the transmission spectrum and the absorption profile.

\section{Observations and Data Reduction}
\label{sec:data}

We observed one transit of the hot Jupiter \object{WASP-52b} on the night of August 28, 2015, using the Optical System for Imaging and low-Intermediate-Resolution Integrated Spectroscopy \citep[OSIRIS;][]{2012SPIE.8446E..4TS} mounted at the Nasmyth-B focal station of the GTC. 

The observation was carried out in the long-slit mode with the R1000R grism and a 40$''$-wide slit. The R1000R grism covers a wavelength range of 515--1025~nm with an instrumental dispersion of $\sim$2.6~\AA\ per pixel. The time-series data were collected by two red-optimized 2048$\times$4096 Marconi CCDs in the standard 200~kHz and 2$\times$2 binning readout mode without windowing, which gives a pixel scale of 0.254$''$ and an overhead of 23.5 seconds between two consecutive exposures. A reference star (\object{2MASS 23140726+0839292}; $r'$mag = 11.3) at a separation of 6.5$'$ was monitored simultaneously with \object{WASP-52} ($r'$mag = 11.5), and they were placed on different CCD chips. 

The transit event was observed from 22:43 UT to 01:25 UT. A total of 308 frames were recorded at an exposure time of 7.5 seconds. The weather was clear in the first two hours, but became poor after the egress due to cirrus crossing. The moon was 99\% illuminated and 30$^{\circ}$ away. The airmass monotonically decreased from 1.55 to 1.07. The seeing varied between 0.87$''$ and 1.45$''$ with a median value of 1.06$''$, which was measured as the full width at half maximum (FWHM) of the target spatial profile at the central wavelength. This resulted in a seeing limited spectral resolution of $\sim$10~\AA. For both spatial and dispersion directions, the positions of spectra remained well within 1 pixel. We did not find any significant rotator-angle dependent systematics as reported in previous observations \citep[e.g.,][]{2016A&A...594A..65N,chen2016a,chen2016b}, probably because this observation only covered a small range of rotator angles ($-229^{\circ}$ to $-212^{\circ}$). 

We reduced the OSIRIS data using the approach outlined in \citet{chen2016a,chen2016b}. The one-dimensional spectra (see Fig.~\ref{fig:GTCSpectra}) were extracted using the optimal extraction algorithm \citep{1986PASP...98..609H} with an aperture diameter of 24 pixels, which minimized the scatter of the white-color light curves created from various trial aperture sizes. The time stamp was centered on mid-exposure and converted into the Barycentric dynamical time standard \citep[\bjdtdb;][]{2010PASP..122..935E}. Any misalignment between the target and reference stars, and any spectral drifts were corrected in the wavelength solutions. Then the requested wavelength range of a given passband was converted to a pixel range, and the flux was summed to generate the time series. The white-color light curve was integrated from 515~nm to 905~nm, which excluded the range of 755--765~nm to eliminate the noise introduced by the telluric oxygen-A band. Wavelengths redder than 905~nm were not used due to second order contamination and fringing.  

\section{Analysis}\label{sec:analysis}

\subsection{Light-curve analysis}
\label{sec:lcfit}

The light-curve data were fit by a model that combines both transit \citep{2002ApJ...580L.171M} and systematics in a multiplicative form, as detailed in \citet{chen2016b,chen2016a}. The combined model was parameterized as mid-transit time (\Tmid), orbital inclination ($i$), scaled semi-major axis (\aRs), planet-to-star radius ratio (\RpRs), quadratic limb-darkening coefficients ($u_1$, $u_2$), and coefficients of systematics models ($c_j$). A circular orbit was assumed and the orbital period was fixed. The limb-darkening coefficients were always imposed with Gaussian priors ($u_i\pm0.1$), whose values were calculated using the Python package written by \citet{2015MNRAS.450.1879E} with the stellar parameters ($T_{\rm eff}$, $\log g$, [Fe/H]) from \citet{2013A_A...549A.134H}. A customized version of the Transit Analysis Package \citep[TAP;][]{2012AdAst2012E..30G} was employed to perform the Markov chain Monte Carlo analysis to find the best-fitting parameters and associated uncertainties, which accounted for the correlated noise using a wavelet-based likelihood function \citep{2009ApJ...704...51C}. 

\begin{figure}
\centering
\includegraphics[width=1\linewidth]{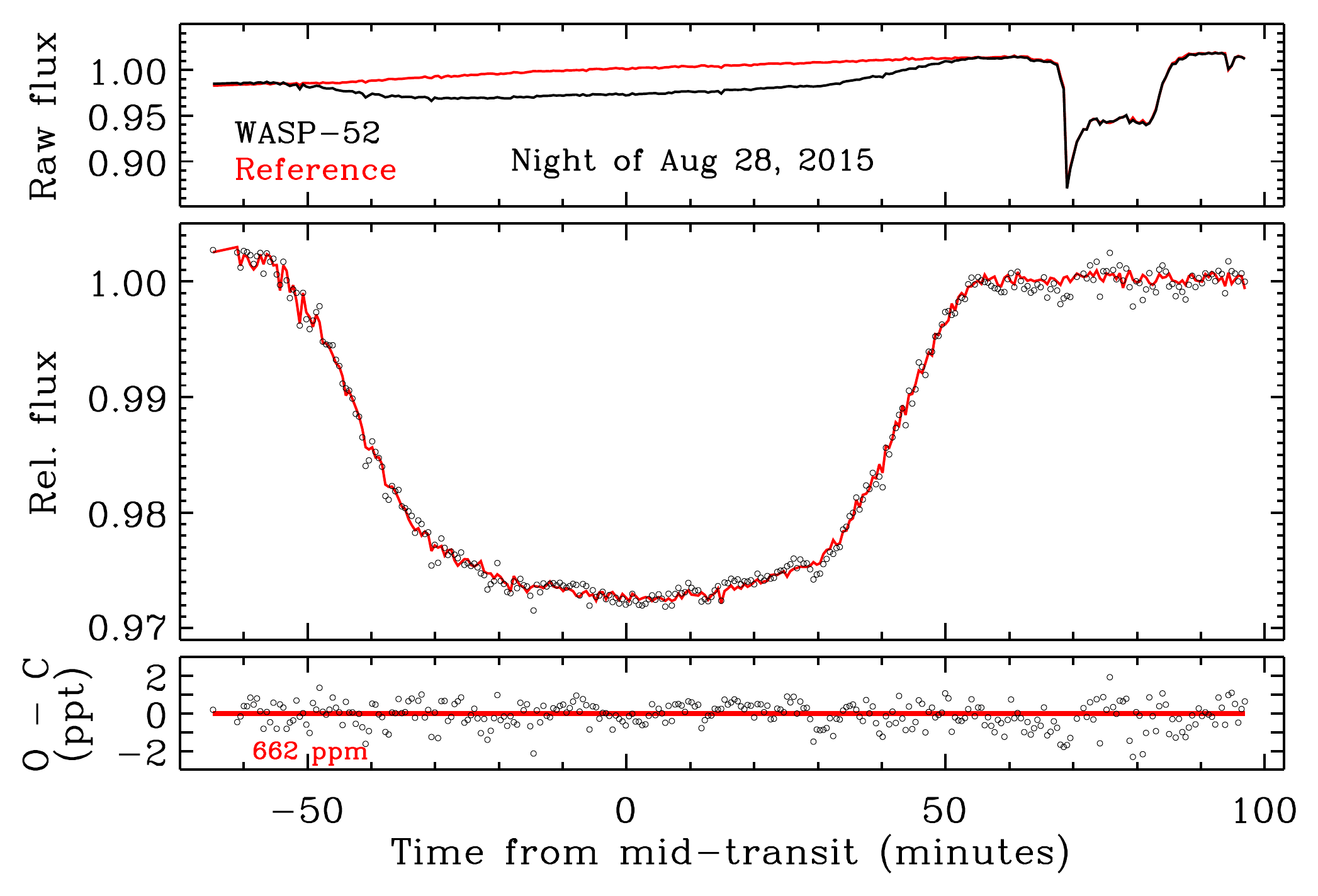}
\caption{Panels from top to bottom: (1) raw flux of WASP-52 (black line) and the reference star (red line) obtained with GTC/OSIRIS; (2) reference-calibrated light curve (black circles) and the best-fitting combined model (red line); (3) best-fitting light-curve residuals. \label{fig:GTCWhiteLC}}
\end{figure}

We first fit the white-color light curve to derive the overall transit parameters (see Table~\ref{tab:tran_param} and Fig.~\ref{fig:GTCWhiteLC}). The systematics model was a linear combination of spatial FWHM and airmass, which was chosen by the Bayesian information criterion \citep[BIC;][]{Schwarz1978}. The resulting transit parameters agree well with previous studies \citep{2013A_A...549A.134H,2016MNRAS.463.2922K,2017MNRAS.465..843M}, except for \RpRs, which appears shallower than previously reported. This deviation is probably caused by stellar activity, given that occulted star spots \citep{2017MNRAS.465..843M} and faculae \citep{2016MNRAS.463.2922K} have been identified  several times. 

\begin{table}
     \small
     \centering
     \caption{System parameters}
     \label{tab:tran_param}
     \begin{tabular}{lr}
     \hline\hline\noalign{\smallskip}
     Parameter & Value\\\noalign{\smallskip}
     \hline\noalign{\smallskip}
     $P$ [days]   & 1.7497798 (fixed) \\\noalign{\smallskip}
     $e$   & 0 (fixed) \\\noalign{\smallskip}
     \Tmid\ [$\mathrm{BJD}_\mathrm{TDB}$] & 2457263.49829 $\pm$ 0.00016\\\noalign{\smallskip}
     $i$ [$^\circ$] & 85.06 $\pm$ 0.27 \\\noalign{\smallskip}
     $a/R_\star$ &  7.14 $\pm$ 0.12 \\\noalign{\smallskip}
     \RpRs & 0.1608 $\pm$ 0.0018 \\\noalign{\smallskip}
     $u_1$ & 0.443 $\pm$ 0.069 \\\noalign{\smallskip}
     $u_2$ & 0.147 $\pm$ 0.091 \\\noalign{\smallskip}
    \hline\noalign{\smallskip}
    \end{tabular}
\end{table}

We then fit the spectroscopic light curves to derive the transmission spectrum. A common-mode trend was removed from all the spectroscopic light curves, which was derived after dividing the white-color light curve by the best-fitting transit model. The remaining wavelength dependent effects were characterized by a linear trend of time. The values of ($i$, \aRs, \Tmid) were fixed to the ones listed in Table~\ref{tab:tran_param}. Two sets of transmission spectra were derived: (1) 22 bins of width 16.5~nm outside of Na and K, 3 bins of 16~\AA\ centered on Na and K, and 1 bin of 18~\AA\ in-between the K doublet; (2) regions outside of Na and K also divided into bins of width 16~\AA. Set 2 provides uniform narrow bins as a double check for Set 1. We present Set 1 spectroscopic light curves in Fig.~\ref{fig:GTCSpecLC} and their transit depths in Table~\ref{tab:gtc_transpec} in the Appendix. The standard deviation of white and spectroscopic light-curve residuals were typically 3.3$\times$ and 1.00--1.33$\times$ photon noise, respectively.

The transit depth at the Na line ($\RpRs=0.1717\pm 0.0030$) is significantly larger than the rest of the 16+18~\AA\ transit depths, which have a nearly Gaussian distribution around $\RpRs=0.1608$. We show the H$\alpha$ line as a control example in Fig.~\ref{fig:GTCTranspec}. Excluding the transit depths of Na and K, we fit a horizontal line to the continuum to estimate the significance of deviation, and found 3.6$\sigma$ for Na, 2.2$\sigma$ for K D$_2$, 0.9$\sigma$ for K D$_1$, and 2.2$\sigma$ for K D$_2$+D$_1$ combined. 

\subsection{Integrated absorption depth}

Following the approach of \citet{2002ApJ...568..377C}, we derived the integrated absorption depth (AD) for the Na line, similar to the methodology in e.g., \citet{2008ApJ...686..658S,2011A&A...527A.110V,2012MNRAS.422.2477H}. We created reference-calibrated light curves for one band centered on Na (5893~\AA) and two bands bracketing this central band (5818--5843~\AA\ and 5943--5968~\AA). Each spectroscopic light curve was modeled in the same way as outlined in Sect.~\ref{sec:lcfit}. After removing the systematics trends and limb-darkening effect, we divided the corrected central-band light curve by the mean of the corrected bracketing-band light curves, and derived the AD in the same way as \citet{2002ApJ...568..377C}. The central band has bin widths varying from 16~\AA\ to 88~\AA\ in a step of 4~\AA. We did not use bin widths narrower than 16~\AA, which would be below the seeing-limited spectral resolution. As shown in Fig.~\ref{fig:NaKADgrowth}, the integrated Na AD drops quickly with increasing bin widths until $\sim$36~\AA, and then slowly approaches a flat continuum, where the largest AD is in the 16~\AA\ bin ($\Delta F/F=0.00378\pm 0.00068$). This indicates that the core of the Na line is very narrow. 

We also derived the AD for the K doublet (D$_2$: 7665~\AA; D$_1$: 7699~\AA) individually. Due to the sharp and strong oxygen-A band, we were not able to construct the bracketing bands directly neighboring to the central band, which were instead set as 7462--7512~\AA\ and 7852--7902~\AA. Since the K D$_1$ line is almost indistinguishable from the neighboring wavelengths (see Fig.~\ref{fig:GTCTranspec}), its AD is heavily contaminated by the neighbors at larger bin widths. On the other hand, the K D$_2$ line, which was 2.2$\sigma$ higher than the continuum, shows a similar AD profile to the Na line (see Fig.~\ref{fig:NaKADgrowth}), where the largest AD is also in the 16~\AA\ bin ($\Delta F/F=0.00149\pm 0.00049$). 

\section{Discussion}
\label{sec:discuss}

\subsection{Transmission spectrum}
\label{sec:transpec}

\begin{figure}
\centering
\includegraphics[width=\linewidth]{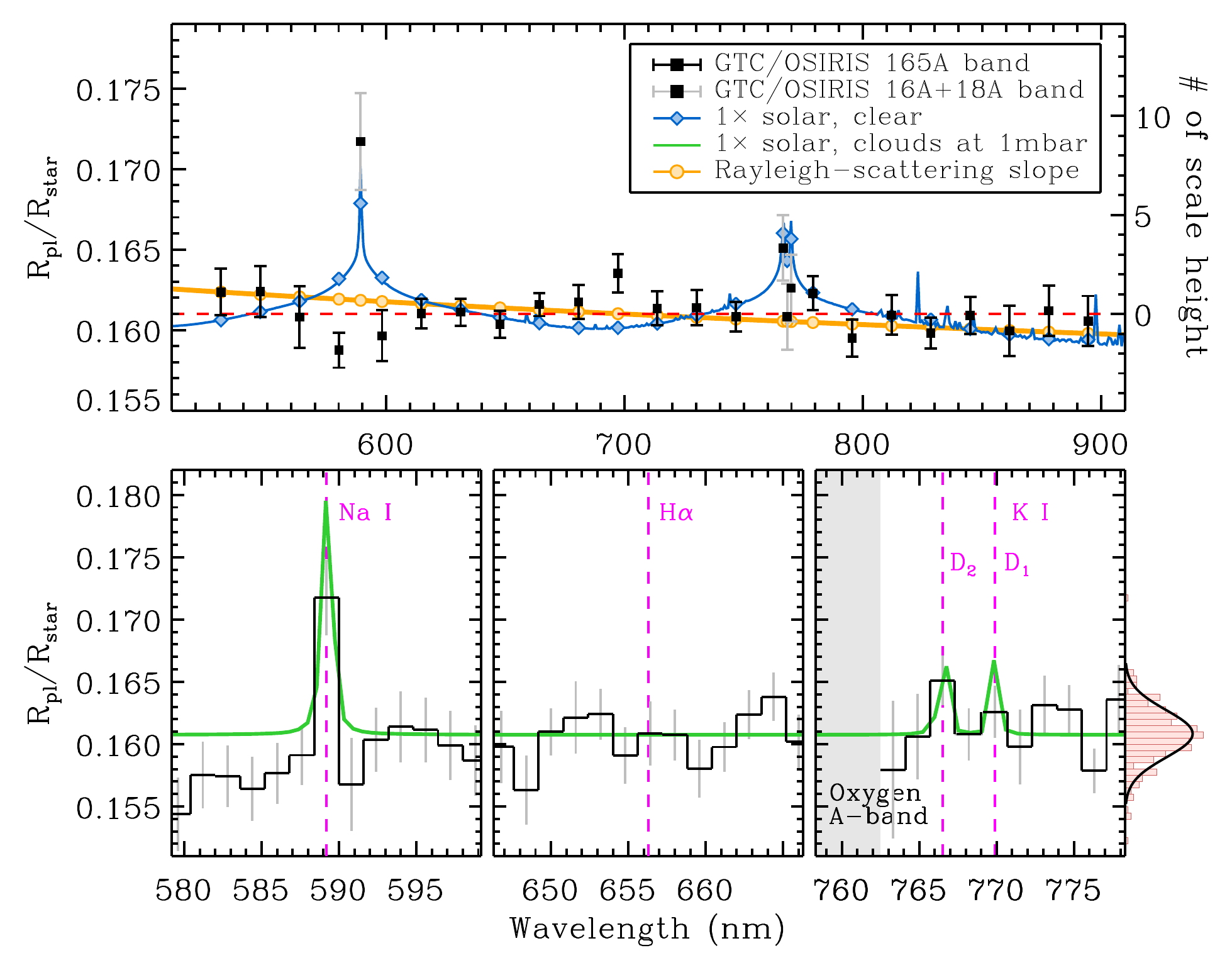}
\caption{GTC/OSIRIS transmission spectrum of \object{WASP-52b}. Top panel: the transmission spectrum composed of 22 bins of 16.5~nm (black squares with black error bars), 3 bins of 16~\AA\ centered on Na and K, and 1 bin of 18~\AA\ in-between the K doublet (black squares with gray error bars). The blue line shows a 1300~K 1$\times$ solar cloud-free model \citep{2016arXiv161103871K}. The orange line shows a Rayleigh-scattering model ($\alpha=-4$). The red dashed line shows a flat line at $\RpRs=0.1608$. Bottom panels: the close-up of Na, H$\alpha$, and K lines (black histogram with gray error bars) in bins of 16~\AA\ (or 18~\AA). The green line shows a 2700~K 1$\times$ solar atmosphere with the clouds at 1~mbar. The red histogram on the right shows the distribution of all the 16+18~\AA\ transit depths.\label{fig:GTCTranspec}}
\end{figure}

As shown in Fig.~\ref{fig:GTCTranspec}, the derived GTC/OSIRIS transmission spectrum for \object{WASP-52b} does not exhibit any broad spectral signatures, except for the narrow excess absorption at the Na line. The standard deviation of the 16.5~nm transit depths $\sigma(\RpRs)=0.00110$ is smaller than one atmospheric scale height $H/R_\star=kT/(\mu gR_\star)=0.00126$, where the stellar radius $R_\star=0.786~R_\sun$, the planetary surface gravity $g=6.85$~m~s$^{-2}$ and the planetary equilibrium temperature $T_\mathrm{eq}=1315$~K were taken from \citet{2017MNRAS.465..843M}; $k$ is the Boltzmann constant; $\mu$ is the mean molecular weight (MMW) and assumed as 2.3$\times$ proton mass. 

We fit a horizontal line with a constant value and a line with a slope in the ($\ln\lambda$, \RpRs) space to the 16.5~nm transit depths, corresponding to (i) clouds with particles comparable to or larger than the probed wavelengths and (ii) clouds with small particles, respectively. The horizontal fit resulted in a slightly larger chi-square value, but it had a $\Delta\mathrm{BIC}=2.9$ lower than the slope fit. For the slope fit, we obtained 
\begin{equation}
\alpha T=\frac{\mu g}{k}\frac{\mathrm{d}R_\mathrm{p}}{\mathrm{d}\ln\lambda}=-873\pm 1824~\mathrm{K},
\end{equation}
where $\alpha$ is the index of the power law cross-section $\sigma=\sigma_0(\lambda/\lambda_0)^\alpha$ for the atmospheric opacity sources \citep{2008A&A...481L..83L} and $\alpha=-4$ corresponds to Rayleigh scattering. With $T=T_\mathrm{eq}$ and $\mu=2.3m_\mathrm{H}$, we calculated $\alpha=-0.7\pm 1.4$, which deviates from $-4$ at 2.4$\sigma$ and indicates that the cloud particles are not likely small.

To distinguish between clear and cloudy scenarios, we also computed isothermal atmospheric models of different temperatures and metallicities with clouds at different pressure levels using the \texttt{Exo-Transmit} code \citep{2016arXiv161103871K}, which were compared to our transmission spectrum (22 bins of 16.5~nm and 4 bins of 16+18~\AA). Atmospheric models with clouds at the pressure of 1~mbar ranked higher in the $\chi^2$ statistics. In Fig.~\ref{fig:GTCTranspec}, we show the 1$\times$ solar cloud-free model in the top panel, and show one of the best atmospheric models (2700~K, 1$\times$ solar, clouds at 1~mbar) in the bottom panels. These two models have a reduced chi-square of $\chi^2_{\rm r}=2.09$ and 0.79 ($d.o.f.=25$), respectively. For comparison, the fit of a horizontal line results in $\chi^2_{\rm r}=1.42$. 

We calculated the cloudiness index $C$ following the approach of \citet{2016ApJ...826L..16H}, which was based on measuring the transit radii at the line center and wing of the Na/K lines, with the cloud-free atmosphere being at $C=1$. For \object{WASP-52b}, the cloudiness index is $C_\mathrm{Na}=5.3\pm 3.0$ based on the Na line, and $C_\mathrm{K}=6.9\pm 3.8$ based on the K D$_2$ line. We used the 28 and 16~\AA\ bins for Na and K, respectively, which enables us to compare our results with \object{HAT-P-1b} \citep[$C_\mathrm{Na}=5.5\pm 2.7$, $C_\mathrm{K}=8.5\pm 7.9$;][]{2016ApJ...826L..16H}. These two planets have almost the same equilibrium temperature, mass, radius, surface gravity, and happen to have the same cloudiness index, but the host star \object{WASP-52} (K2V) is far more active than \object{HAT-P-1} (G0V), making them good comparative targets for the origin of clouds.  

We note that in principle the best-fitting slope in the transmission spectrum can be introduced by unocculted star spots or occulted faculae, which would indicate shallower true transit depths than currently measured. However, the correlated noise in our white-color light curve, which might arise from occulted star spots or occulted faculae, does not show any dependency of wavelength (see Fig.~\ref{fig:GTCSpecLC}). Furthermore, we do not observe any excess or deficient absorption in other stellar lines (e.g., H$\alpha$, \ion{Ca}{II}~IRT). Therefore, we argue that the excess Na absorption is not likely a contamination of stellar activity.

\subsection{Thermal structure for the upper atmosphere}

\begin{figure}
\centering
\includegraphics[width=\linewidth]{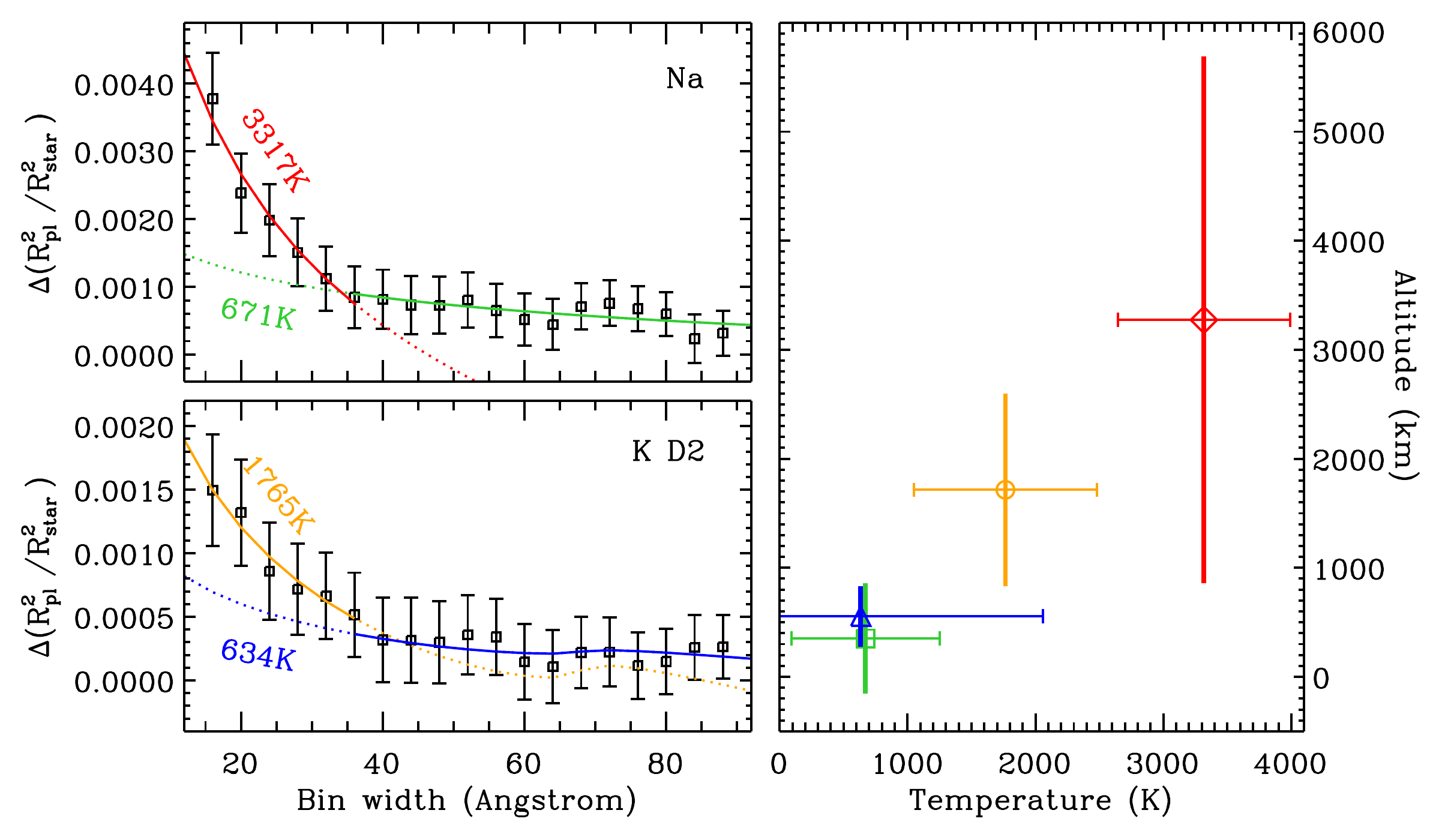}
\caption{The left panels show the absorption depths integrated in different bin widths for Na (top) and K D$_2$ (bottom). The red/orange (green/blue) curves present the best-fitting isothermal models for the core (wing) region. The right panel shows the temperatures of these isothermal models at corresponding altitude ranges.\label{fig:NaKADgrowth}}
\end{figure}

\citet{2011A&A...527A.110V,2011A&A...533C...4V} and \citet{2012MNRAS.422.2477H} have derived the temperature structures for the upper atmospheres of \object{HD 209458b} and \object{HD 189733b} by investigating the Na AD profile. We followed the approach outlined in \citet{2012MNRAS.422.2477H} to generate isothermal models for the Na and K lines. The model was binned in the same way in which we created the AD profile. Given the discussion in Sect.~\ref{sec:transpec}, we assumed the pressure at the reference level to be 1~mbar, the Na and K abundances to be 1$\times$ solar, and the planet radius to be $\RpRs=0.1608$ (i.e., the continuum altitude). Assuming that the probed altitude range is isothermal, the slope of the AD profile is a function of temperature, similar to the case of Rayleigh scattering. 

We fit the model to the bin width ranges of 16--36\AA\ and 40--88\AA, corresponding to the line core and wings, respectively. The results are shown in Fig.~\ref{fig:NaKADgrowth}. The temperatures inferred from the line wings of Na and K are almost the same ($T\approx665$~K), while the Na line core has a much higher temperature ($T=3320\pm 670$~K) than the K D$_2$ core ($T=1765\pm 715$~K) as a result of a wider and higher probed altitude range. However, it is possible that the Na core temperature is overestimated due to decreasing of MMW at higher altitude \citep[e.g.,][]{2004Icar..170..167Y,2007P&SS...55.1426G}. Assuming a constant $\mu=2.3$~$m_\mathrm{H}$ over all altitudes, we estimated that the Na line probed a pressure range of 1500--87~$\mu$bar at the wing and 87--5~$\mu$bar at the core, and the K D$_2$ line probed 430--82~mbar at the wing and 82--12~$\mu$bar at the core, individually. Therefore, the inferred average temperature-pressure profile of \object{WASP-52b} shows an increasing trend towards higher upper atmosphere, with a positive temperature gradient of $0.88\pm 0.65$~K\,km$^{-1}$, which is similar to \object{HD 209458b} \citep{2011A&A...527A.110V,2011A&A...533C...4V}, \object{HD 189733b} \citep{2012MNRAS.422.2477H,2015ApJ...803L...9H}, and \object{WASP-49b} \citep{2017arXiv170200448W}, and might be indicative of upper atmospheric heating and escaping processes  \citep[e.g.,][]{2003ApJ...598L.121L,2004A&A...418L...1L,2013Icar..226.1678K,2013Icar..226.1695K}. This also suggests \object{WASP-52b} as a potential candidate of evaporating planets \citep[e.g.,][]{2003Natur.422..143V,2010A&A...514A..72L,2015Natur.522..459E}.

On the other hand, as pointed by \citet{2012MNRAS.422.2477H}, the integrated AD profile could be compromised by the dilution effects of increasing bandwidth around an unresolved line. Therefore, future observations of high-resolution spectroscopy are required to confirm this line-profile diagnosis, which could also provide useful information on the impact of stellar activity.

\section{Conclusions}\label{sec:conclusions}

We have observed one transit of \object{WASP-52b} with GTC/OSIRIS. The resulting transmission spectrum does not show any broad spectral signature except for a prominent and narrow excess absorption at the Na line, indicative of a cloudy atmosphere. The integrated absorption depth profile of Na and K might suggest an inverted temperature structure for the upper atmosphere. Future ground-based high-resolution spectroscopy are required to put further constraints on line-profile diagnosis and stellar activity. 

\begin{acknowledgements}
    The authors thank the referee K. Heng for the useful and constructive comments.
    This research is based on observations made with the Gran Telescopio Canarias (GTC), installed in the Spanish Observatorio del Roque de los Muchachos, operated on the island of La Palma by the Instituto de Astrof\'isica de Canarias.  
    This work is partly financed by the Spanish Ministry of Economics and Competitiveness through projects ESP2013-48391-C4-2-R, ESP2014-57495-C2-1-R, and AYA2012-39346-C02-02. 
    G.C. also acknowledges the support by the National Natural Science Foundation of China (Grant No. 11503088) and the Natural Science Foundation of Jiangsu Province (Grant No. BK20151051).
    This research has made use of the VizieR catalogue access tool, CDS, Strasbourg, France \citep{2000A&AS..143....9W}.
\end{acknowledgements}

\bibliographystyle{aa} 
\bibliography{ref_db} 


\begin{appendix}
\section{Additional figures and tables.}

\begin{figure}[h!]
\centering
\includegraphics[width=\linewidth]{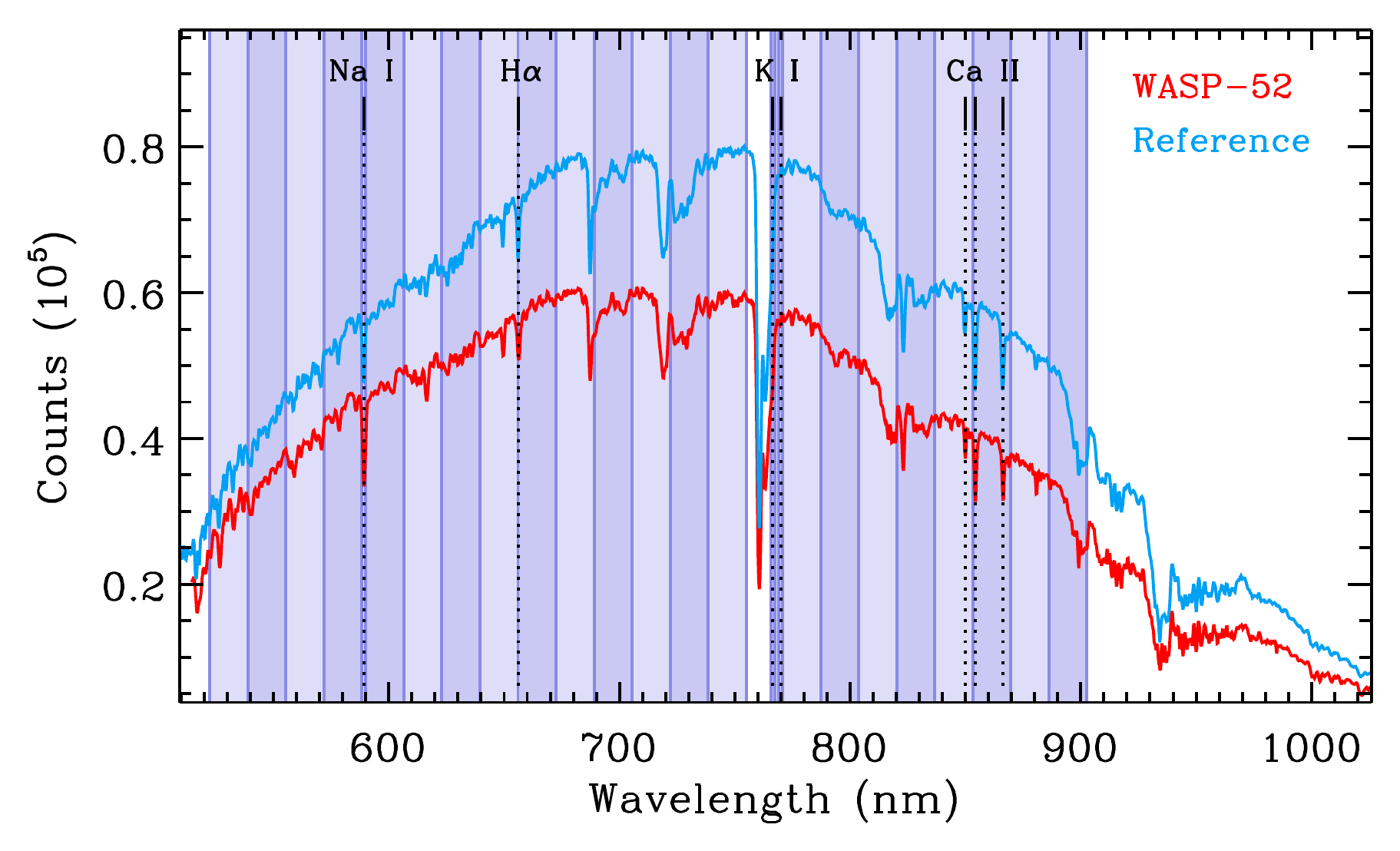}
\caption{Example stellar spectra of \object{WASP-52} (red) and the reference star (blue) obtained with the R1000R grism of GTC/OSIRIS on the night of August 28, 2015. The color-shaded areas indicate the divided passbands that are used to create the spectroscopic light curves.\label{fig:GTCSpectra}}
\end{figure}

\begin{figure*}[h!]
\centering
\includegraphics[width=0.88\linewidth]{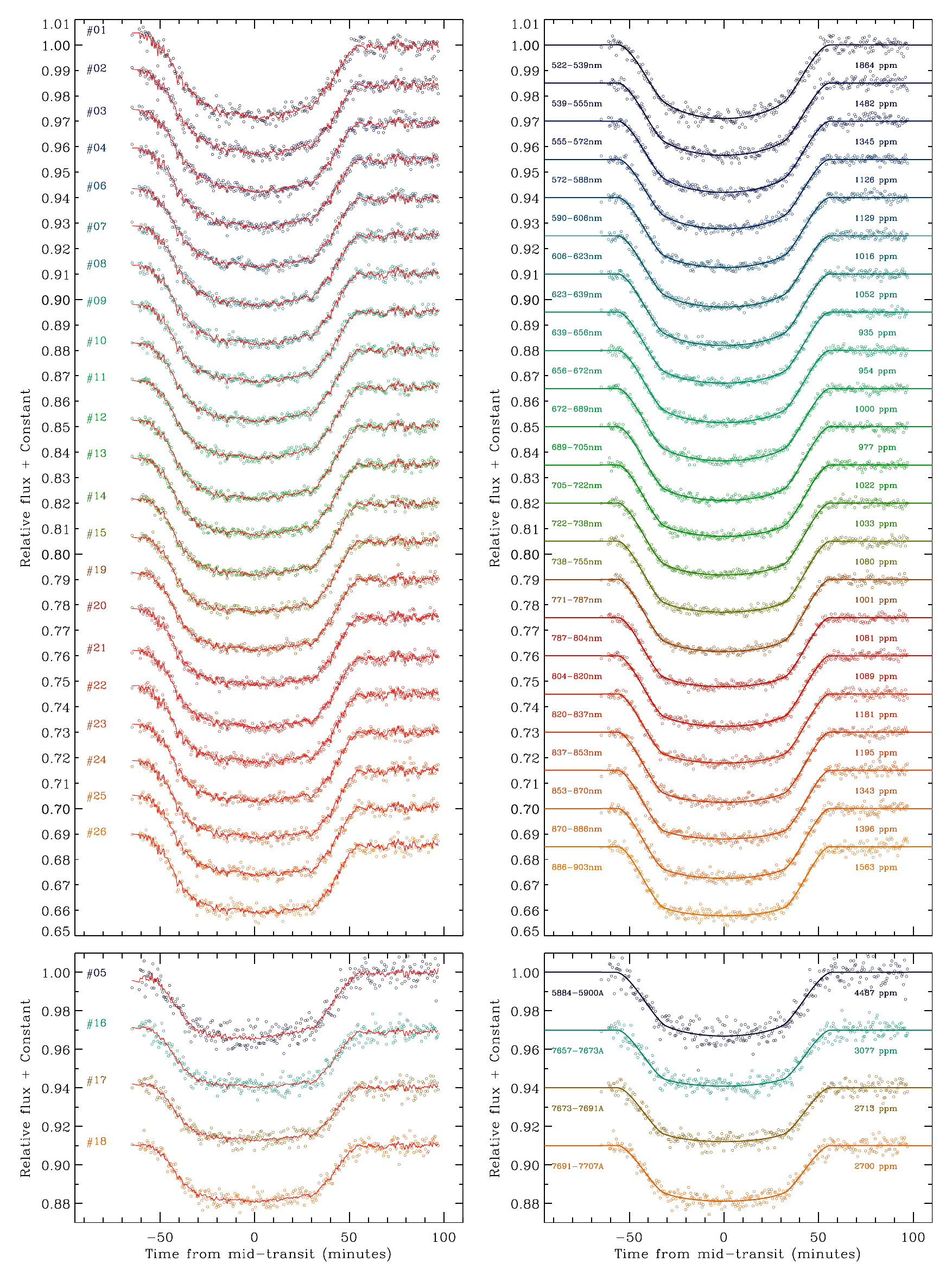}
\caption{Raw (left panel) and detrended (right panel) spectroscopic light curves of WASP-52b obtained with the R1000R grism of GTC/OSIRIS, including 22 channels of 16.5~nm, 3 channels of 16~\AA, and 1 channel of 18~\AA, as indicated in Fig.~\ref{fig:GTCSpectra}. For clarity, the four narrow-band light curves are shown separately in the bottom panels.\label{fig:GTCSpecLC}}
\end{figure*}

In this appendix, we show the example stellar spectra in Fig.~\ref{fig:GTCSpectra}, the spectroscopic light curves in Fig.~\ref{fig:GTCSpecLC}, and the derived wavelength dependent transit depths in Table~\ref{tab:gtc_transpec}.

\begin{table}[h!]
     \footnotesize
     \centering
     \caption{Transmission spectrum obtained with GTC/OSIRIS}
     \label{tab:gtc_transpec}
     \begin{tabular}{cccc}
     \hline\hline\noalign{\smallskip}
     \# & \multicolumn{2}{c}{Wavelength~(\AA)} & \RpRs \\\noalign{\smallskip}
        & Center & Width&  \\\noalign{\smallskip}
     \hline\noalign{\smallskip}
 1 &   5306 &    165 & 0.1624 $\pm$ 0.0015 \\\noalign{\smallskip}
 2 &   5471 &    165 & 0.1624 $\pm$ 0.0016 \\\noalign{\smallskip}
 3 &   5636 &    165 & 0.1608 $\pm$ 0.0019 \\\noalign{\smallskip}
 4 &   5800 &    165 & 0.1588 $\pm$ 0.0011 \\\noalign{\smallskip}
 5 &   5891 &     16 & 0.1717 $\pm$ 0.0030 \\\noalign{\smallskip}
 6 &   5981 &    165 & 0.1596 $\pm$ 0.0016 \\\noalign{\smallskip}
 7 &   6147 &    165 & 0.1610 $\pm$ 0.0009 \\\noalign{\smallskip}
 8 &   6312 &    165 & 0.1611 $\pm$ 0.0008 \\\noalign{\smallskip}
 9 &   6477 &    165 & 0.1604 $\pm$ 0.0008 \\\noalign{\smallskip}
10 &   6642 &    165 & 0.1616 $\pm$ 0.0007 \\\noalign{\smallskip}
11 &   6807 &    165 & 0.1617 $\pm$ 0.0011 \\\noalign{\smallskip}
12 &   6972 &    165 & 0.1635 $\pm$ 0.0012 \\\noalign{\smallskip}
13 &   7137 &    165 & 0.1613 $\pm$ 0.0011 \\\noalign{\smallskip}
14 &   7302 &    165 & 0.1614 $\pm$ 0.0011 \\\noalign{\smallskip}
15 &   7467 &    165 & 0.1608 $\pm$ 0.0009 \\\noalign{\smallskip}
16 &   7665 &     16 & 0.1651 $\pm$ 0.0020 \\\noalign{\smallskip}
17 &   7682 &     18 & 0.1608 $\pm$ 0.0021 \\\noalign{\smallskip}
18 &   7699 &     16 & 0.1626 $\pm$ 0.0021 \\\noalign{\smallskip}
19 &   7790 &    165 & 0.1623 $\pm$ 0.0011 \\\noalign{\smallskip}
20 &   7955 &    165 & 0.1595 $\pm$ 0.0011 \\\noalign{\smallskip}
21 &   8120 &    165 & 0.1609 $\pm$ 0.0012 \\\noalign{\smallskip}
22 &   8285 &    165 & 0.1598 $\pm$ 0.0010 \\\noalign{\smallskip}
23 &   8450 &    165 & 0.1609 $\pm$ 0.0012 \\\noalign{\smallskip}
24 &   8615 &    165 & 0.1599 $\pm$ 0.0016 \\\noalign{\smallskip}
25 &   8780 &    165 & 0.1612 $\pm$ 0.0016 \\\noalign{\smallskip}
26 &   8945 &    165 & 0.1606 $\pm$ 0.0016 \\\noalign{\smallskip}
    \hline\noalign{\smallskip}
    \end{tabular}
\end{table}

\vspace{10em}

\end{appendix}

\end{document}